\documentclass[twocolumn,showpacs,preprintnumbers,amsmath,amssymb]{revtex4}

\usepackage{graphicx}
\usepackage{amssymb}
\usepackage{hyperref}

\begin{document}
\title{Suppression of the Shastry-Sutherland phase in Ce$_{2}$Pd$_{2}$Sn at a field induced critical point}

\author{J.G. Sereni$^{1}$, M. G\'omez Berisso$^{1}$, G. Schmerber$^2$, A. Braghta$^{2,3}$ and J.P. Kappler$^2$}
\address{$^1$ Div. Bajas Temperaturas, Centro At\'omico Bariloche (CNEA), 8400 S.C. Bariloche, Argentina\\
$^2$ IPCMS, UMR 7504 CNRS-ULP, 23 rue de Loess, B.P. 43 Strasbourg
Cedex 2, France\\$^3$ D\'epartement de Physique, Universit\'e de
Guelma, 24000 Guelma, Algeria}

\date{\today}

\begin{abstract}

{The magnetic phase diagram of Ce$_2$Pd$_2$Sn is investigated
through the field dependence of thermal, transport and magnetic
measurements performed at low temperature. The upper transition,
$T_M=4.8$\,K is practically not affected by magnetic field up to
$B=1$\,T, whereas the lower one $T_C(B)$ rapidly increases from
2.1\,K joining $T_M$ in a critical point at $T_{cr}=(4.2\pm 0.3$)K
for $B_{cr}=(0.12\pm 0.03)$\,T. At that point the intermediate
phase, previously described as an unstable Shastry-Sutherland
phase, is suppressed. A detailed analysis around the critical
point reveals a structure in the maximum of the $\partial
M/\partial B(B)$ derivative which could be related to the
formation of a novel phase in that critical region. }

\end{abstract}

 \maketitle

%\keywords{Keywords: Cerium Intermetallics, Critical Points,
%Magnetic Phase Diagrams}
%\end{frontmatter}
\section{Introduction}

Complex or metastable phases are favored in the proximity of
magnetic transitions since, within the Landau theory, the change
of sign of the first term of the free energy $G(\psi)$ broadens
its minimum as a function of the order parameter $\psi$ as $T\to
T_C$. In these critical conditions the 'roughness' of $G(\psi)$,
occurring in real systems, may present relative minima which may
become relevant in the formation of novel phases \cite{Kirchpat}.
In recent years an increasing number of magnetic systems
exhibiting non trivial types of order parameters have been
reported. Those complex phases may correspond to metastable ground
states, such as frustrated or 'spin ice' \cite{spinice} systems,
Shastry-Sutherland phases \cite{Shastry,Miyahara} or other exotic
phases. Those phases can be tuned by non-thermal control
parameters like pressure (e.g. in non conventional superconductors
\cite{Macknezie}) or by magnetic field (e.g. in Sr$_3$Ru$_2$O$_7$
\cite{Green}) and driven to a quantum critical points at $T=0$
\cite{HVL}. Alternatively, some of those phases become unstable
under moderate variation of the external parameters and are
suppressed in a critical point at finite temperature.

Recently, the formation of an unstable Shastry-Sutherland ShSu
phase was reported to form in Ce$_2$Pd$_2$Sn \cite{Ce2Pd2Sn}. That
phase is observed within a limited range of temperature, between
$T_M=4.8> T > T_C=2.1$\,K, having as the upper limit a correlated
paramagnetic phase and a ferromagnetic FM one as ground state GS.
This exotic phase builds up from FM-dimers formed by Ce
nearest-neighbor atoms, and shows the onset of magnetic
correlations at $T\leq 20$\,K. An antiferromagnetic AF exchange
interaction between those dimers drives the formation of the ShSu
phase, realized in this compound as a quasi 2D square lattice of
effective spin $S_{eff}=1$ below $T_M=4.8$\,K \cite{Ce2Pd2Sn}.

Stable ShSu phases are predicted for AF-dimers and AF inter-dimer
interactions by theoretical calculations \cite{Miyahara}. This
scenario was recently observed in Yb$_2$Pt$_2$Pb single crystals
\cite{Kim}, which undergoes a slight shift of part of Pt atoms
that results in two types of Yb-Pt tetrahedral sublattices. On the
contrary, Ce$_2$Pd$_2$Sn presents a unique crystalline Ce-lattice
where the mentioned exotic phase becomes unstable at $T_C=2.1$\,K,
undergoing a first order to its FM-GS. Neutron diffraction
experiments \cite{Laffarge96} revealed a modulated character of
this phase, with the local moments pointing in the 'c'
crystallographic direction. The incommensurate propagation vector
$[qx]$ changes from 0.11 (at 4.2\,K) to 0.077 (at 2.8\,K) where it
suddenly drops to $[qx=0]$ once the long range FM order parameter
sets on.

Since the mentioned magnetic phase is unstable and only holds
within a short range of temperature, the application of external
field is expected to produce significant effects including its
eventual suppression in a critical point. To our knowledge there
are no systematic investigations of magnetic phase diagrams for
Ce$_{2}$T$_{2}$X compounds (with T = Ni, Cu, Pd, Rh, Pt and X = Sn
and In) performed yet. This is not a minor point since in the last
years a big effort was done searching for critical points at very
low temperature and magnetic field may fine tune the critical
conditions. Furthermore, the scarce examples for FM critical
points places the Ce$_{2}$T$_{2}$Sn family of compounds as good
candidates for that current topic since it builds up from Ce atoms
disposed in triangular prims mimiking those of CeT \cite{Gsch} FM
compounds.

Preliminary studies on the effect of the magnetic field
\cite{Berisso} on Ce$_{2}$Pd$_{2}$Sn showed that the upper
transition is overcome by the lower one by applying moderate
magnetic filed. In order to investigate the characteristics of
that critical region, we have performed a detailed investigation
of the magnetic field effects on the low temperature thermal,
magnetic and transport properties of this compound. Such a study
allows to assemble a magnetic phase diagram including the critical
point where the intermediate phase is suppressed.

\section{Experimental details and results}

\begin{figure}
\begin{center}
\includegraphics[angle=0, width=0.45 \textwidth] {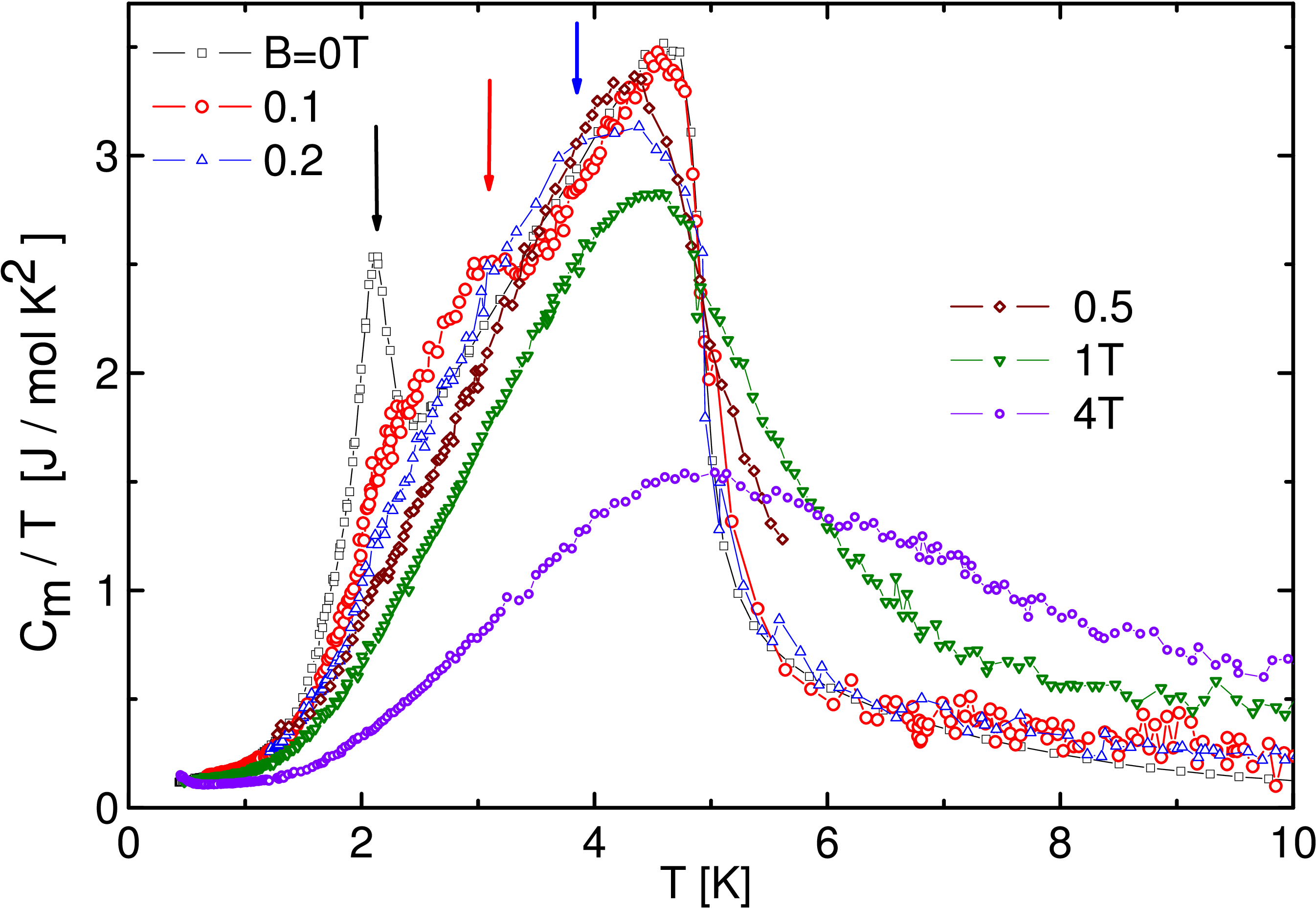}
%{/home/jsereni/papers/publicaz02/pdnial3/textos/F1latpar}
\end{center}
\caption{(Color online) Effect of applied magnetic field up to 4T
on the specific heat after phonon subtraction. Arrows indicate the
field dependence of the lower ($T_C$) transition.} \label{F1}
\end{figure}

Details of sample preparation were described in a previous paper
\cite{Ce2Pd2Sn}. Structural characterization confirms the single
phase composition of the sample in a tetragonal Mo$_2$FeB$_2$-type
structure with $a=7.765\AA$ and $c=3.902\AA$ lattice parameters.
The actual composition of the sample was determined to be
Ce$_{2.005}$Pd$_{1.988}$Sn$_{0.997}$ after SEM/EDAX analysis.

Specific heat was measured using the heat pulse technique in a
semi-adiabatic He-3 calorimeter in the range between 0.5 and 20K,
at zero and applied magnetic field up to 4T. DC-magnetization
measurements were carried out using a standard SQUID magnetometer
operating between 2 and 300K, and as a function of field up to 5T.
Electrical resistivity was measured between 0.5K and room
temperature using a standard four probe technique with an LR700
bridge.

As shown in Fig.~\ref{F1}, the magnetic field affects quite
differently the specific heat $C_m(T)$ anomalies of both
transitions. While the temperature of the lowest ($T_C=2.1$\,K)
rapidly increases with field, the upper one ($T_M=4.8$\,K) reminds
practically unaffected up to $B=0.2$T. The well defined first
order transition at $T_C(B=0)$ transforms into a broaden anomaly
by the effect of applied field, which extends between $2\leq T
\leq 3$\,K for $B=0.1$\,T, see Fig.~\ref{F1}. Such a broadening
can be explained by the poly-crystalline nature of the sample
since the crystals are randomly oriented between the easy and hard
directions of magnetization. This random orientation yields to a
continuous distribution of transitions according to the respective
projections of the crystals respect to the magnetic field.

\begin{figure}
\begin{center}
\includegraphics[angle=0,width= 0.45 \textwidth] {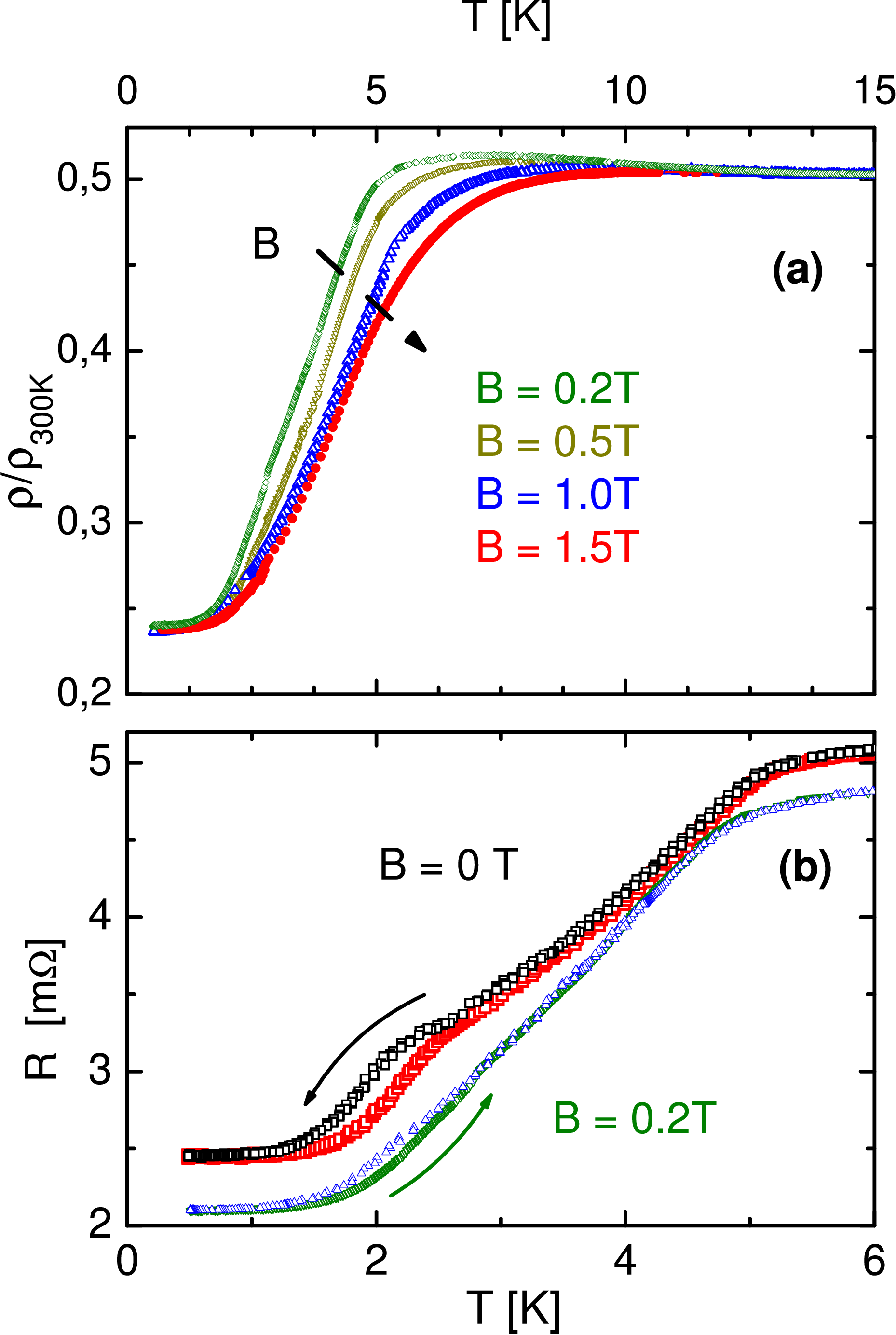}
%{/home/jsereni/papers/publicaz02/pdnial3/textos/F1latpar}
\end{center}
\caption{(Color online) Effect of applied magnetic field on the
electrical resistivity. (a) involving both magnetic transitions
and (b) showing the hysteresis of the lower ($T_C$) first order
transition.} \label{F2}
\end{figure}

The fact that the jump of the specific heat remains unaffected and
the temperature of the maximum value of $C_m$ (at $T=T_M$) does
not decrease up to $B=0.2$\,T are further evidences that the upper
transition cannot be considered as AF in a canonical sense.
Nevertheless, the cusp of $M(T)$ at that temperature reveals that
some type of AF interactions play some role in the magnetic
structure below $T_M$. The lack of variation of $C_m(T)$ for
fields $B\leq 0.2$\,T indicates that the arguments used in
Ref.\cite{Ce2Pd2Sn} to recognize the formation of Ce-Ce pairs
remains valid at low field. At higher field (i.e. $B\geq 0.5T$)
the maximum value of $C_m(T_M)$ starts to decrease with a
concomitant broadening of the anomaly, see Fig.~\ref{F1}. Such a
behavior indicates that an induced ferromagnetic state sets in,
which at $B=4$T clearly tends to a Schottky-type anomaly.

According to the $C_m(T)/T$ behavior depicted in Fig.~\ref{F1},
the temperature dependence of the magnetic entropy gain $S_m$
reduces its thermal increase as the external field increases.
Independently of that reduction, $S_m(T)$ reaches the full
expected value of 2RLn2 at $T\approx 20$\,K even for $B=1T$. For
highest measured field (c.f. $B=4$\,T) only 80\% of the total
entropy of the doublet GS is reached at that temperature (c.f. our
upper limit of temperature).

$\rho(T,B)$ measurements, shown in Fig.~\ref{F2}a, indicate that
magnetic correlations above $T_M$ are progressively quenched by
magnetic field and the transition itself is smeared once the
applied field induces the FM behavior. Concerning the lower
transition, the $\rho(T)$ hysteresis is strongly weakened in a
field of 0.2\,T in agreement with the specific heat results, see
Fig.~\ref{F2}b. This means that the first order transition
vanishes above that field.

\begin{figure}
\begin{center}
\includegraphics[angle=0, width= 0.45 \textwidth] {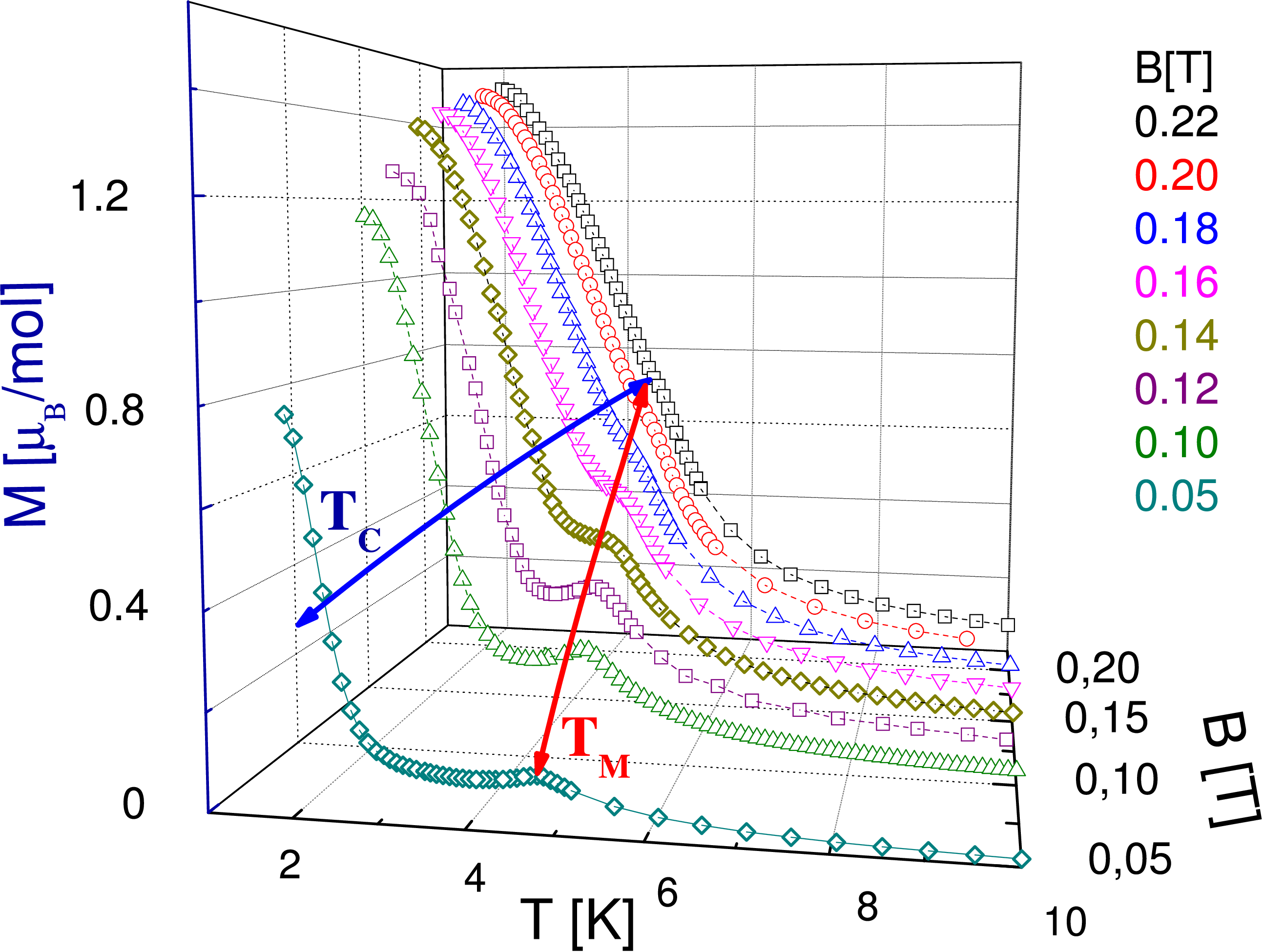}
%{/home/jsereni/papers/publicaz02/pdnial3/textos/F1latpar}
\end{center}
\caption{(Color online) Temperature dependence of the
magnetization for different applied fields, between 50mT and
0.22T. Solid curves the respective $T_C$ and $T_M$ field
dependencies} \label{F3}
\end{figure}

The $M(T)$ dependence of was measured under different applied
fields, starting from $B=0.05$\,T. Between 0.1\,T and 0.22\,T the
isodyna curves $M(T)\mid_B$ were measured with steps of 0.02\,T as
shown in Fig.~\ref{F3}. This study shows how the FM phase
overcomes the modulated one, being the boundary of the former
determined by the maximum slope of $M(T)$ (i.e. $dM/dT\mid_{max}$)
and that of the later by the cusp at $M(T)$ at ($T_M$). Both phase
boundaries are depicted in the phase diagram of Fig.~\ref{F6} by
respective curves which join at a critical point.

Isothermal $M(B)$ measurements were performed within the $2\leq T
\leq 5$\,K range of temperature up to $B=1$\,T. In Fig.~\ref{F4}
we present the low field magnetization curves, with the typical FM
dependence at $T = 2$\,K, including a weak hysteresis (not shown
for clarity) between increasing and decreasing field. Above the
transition (i.e. $T\geq 2.5$\,K), $M(B)$ shows an S-shape
dependence indicating that a re-arrangement of the magnetic
structure occurs as a broad field induced meta-magnetic
transformation. Such a broadening is attributed to the randomly
distribution of the magnetic directions in this poly-crystalline
and strongly anisotropic compound. Interestingly, $M(B)$ curves
start to intersect each other between 4\,K and 4.7\,K, see the
remarked region in the figure around 0.12\,T. This suggests that
some anomalous variation of the magnetization occurs close to the
critical field at that temperature range.

\section{Discussion}

A simple thermodynamic analysis of the field effect on the thermal
dependence of the specific heat allows to extract some trends
concerning the magnetic behavior of the system. From
Fig.~\ref{F1}, one sees that at low field ($B\leq 0.2$\,T)
$C_m(B)/T \mid_T$ slightly increases below $T_M$ and consequently
$S_m(B)\mid_T$ computed as $\int C_m(T)/T dT$. From Maxwell
relations, such increase of $S_m(B)\mid_T$ corresponds to the
observed increase of $M(T)$ as $T\to T_M$ since $\partial
S_m/\partial B \mid _T =
\partial M/\partial T \mid_H >0$. On the contrary, at higher
fields (i.e. $B\geq 0.5$\,T) an induced FM behavior is reflected
in a strong $\partial S_m/\partial B \mid _T$ decrease. This
change of behavior reveals that the apparent AF character of the
intermediate phase is overcome by a FM behavior around 0.2\,T.

\begin{figure}
\begin{center}
\includegraphics[angle=0,width= 0.45 \textwidth] {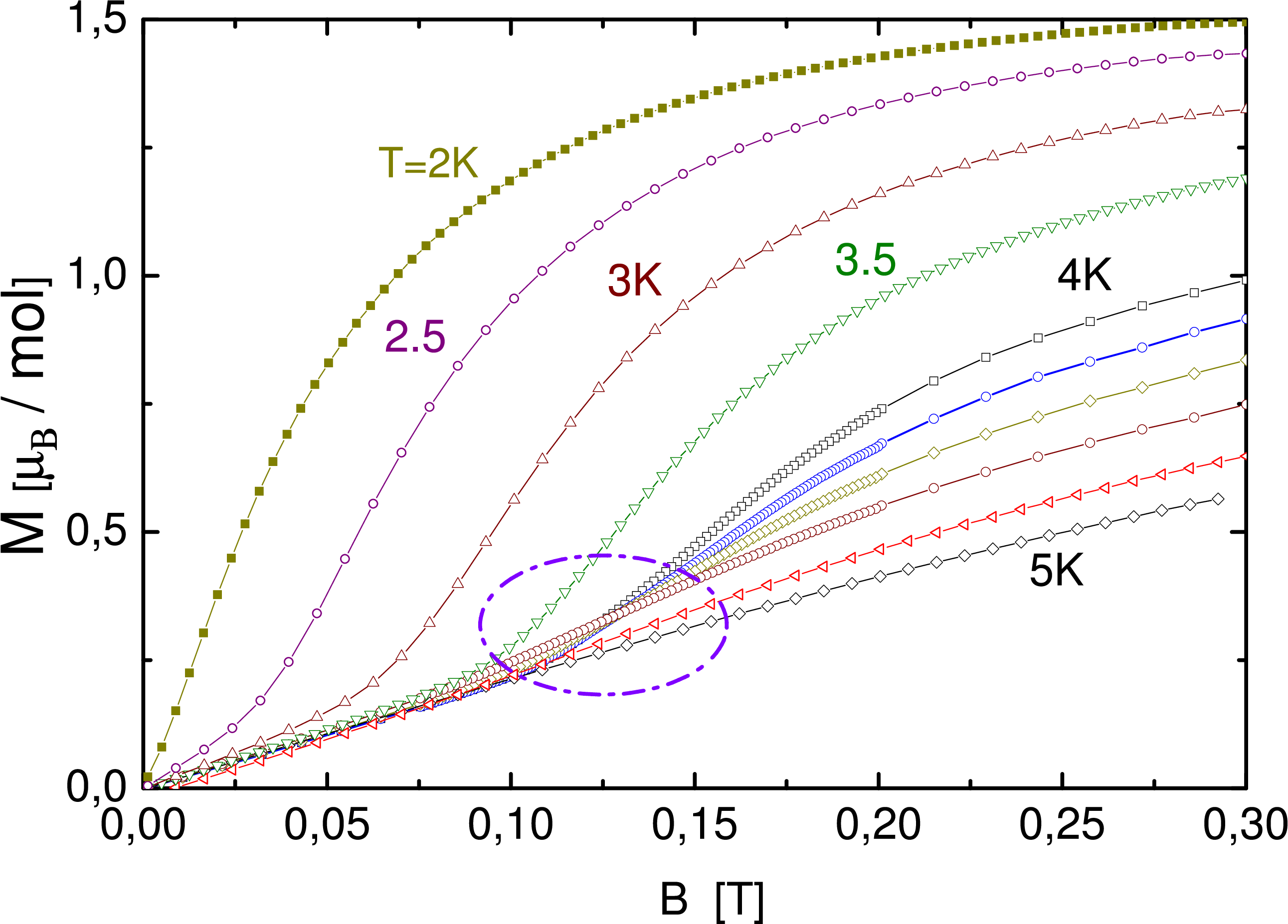}
%{/home/jsereni/papers/publicaz02/pdnial3/textos/F1latpar}
\end{center}
\caption{(Color online) Isothermal field dependence of
magnetization up to 0.3T for the intermediate phase ($T_C<T<T_M$).
Between 4K and 5K the measurements were carried every 0.1K, but
represented every 0.2\,K for clarity. The marked area centered at
at $B=0.12$\,T is discussed in the text.} \label{F4}
\end{figure}

\begin{figure}
\begin{center}
\includegraphics[angle=0,width= 0.45 \textwidth] {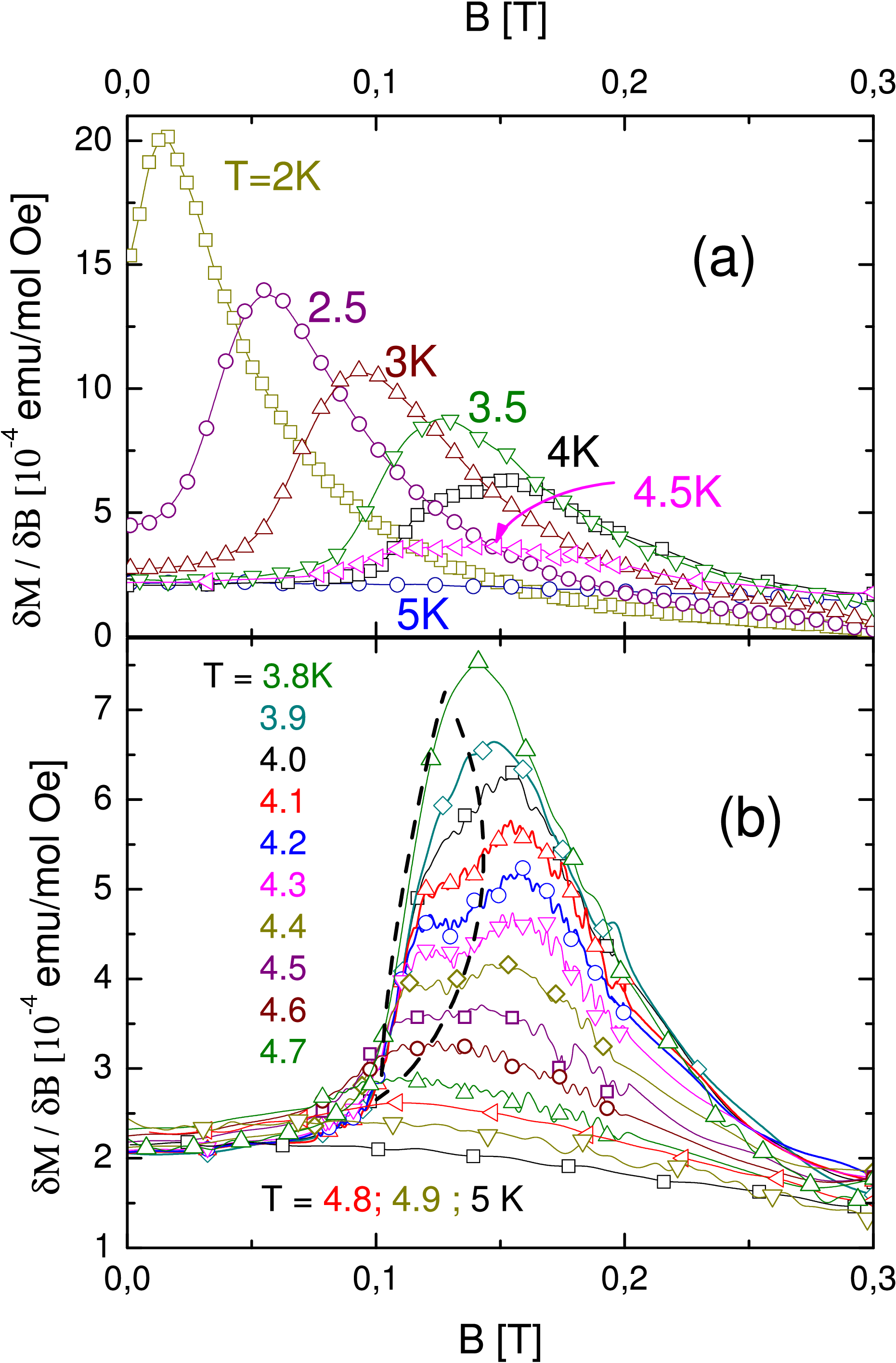}
%{/home/jsereni/papers/publicaz02/pdnial3/textos/F1latpar}
\end{center}
\caption{(Color online) (a) Isothermal field derivative of the
magnetization between 2 and 5\,K and (b) detail between $3.8\leq T
\leq 5$\,K showing an incipient structure in the maximum of the
derivative around the critical point. Dashed curves mark the
maximum slope of those derivatives} \label{F5}
\end{figure}

Since the specific heat jump is progressively smeared by the
mentioned random orientation of individual crystals and the
maximum slope of $\partial M/\partial T\mid_B$ is not properly
defined, to determine the phase boundary between the magnetic
phases for a phase diagram construction requires to use a precise
procedure. For such a purpose we have analyzed the field
dependence of the isothermal $\partial M/\partial B\mid_T$
derivative between 2\,K and 5\,K. From Fig.\ref{F5}a one may
extract the maximum slope of this derivative, which is related
with the maximum curvature of the $M(B)$ dependence (see
Fig.~\ref{F4}) and provides a better approach for determining the
field at which the induced FM phase sets in.

Also in Fig.\ref{F5}a one can see well defined maxima of $\partial
M/\partial B$ isotherms as a function of field, whose intensity
decreases as the temperature increases. Approaching the critical
point, each maximum splits into two maxima hardly distinguish in
the 4.5\,K isotherm. To elucidate the real existence of a
structure within that narrow range of field and temperature, we
have performed detailed measurements of $M(B)$ between 3.8\,K and
4.7\,K up to $B=0.3$\,T.

Those results are presented in Fig.\ref{F5}b. Within that range of
temperature one may distinguish between a main slope of $\partial
M/\partial B\mid_T$, practically fixed at $B\approx 0.11$\,T, and
a second contribution which arises up to $B=0.15$\,T at
$T=4.2$\,K, see dashed curves in Fig.\ref{F5}b. Above that
temperature the satellite maximum progressively vanishes,
disappearing at $T\approx 4.7$\,K. Such a split of the $\partial
M/\partial B\mid_T$ derivative might be due to the formation of a
field induced phase when the point is approached. This feature my
be related to one of the predictions for the ShSu lattice, that is
the appearance of a plateaux in the magnetization at 1/4, 1/8 and
1/10 of the full saturated moment, as different ordered states
emerge from the initially frustrated spin liquid through the
application of magnetic field \cite{Miyahara}.

\begin{figure}
\begin{center}
\includegraphics[angle=0,width= 0.45 \textwidth] {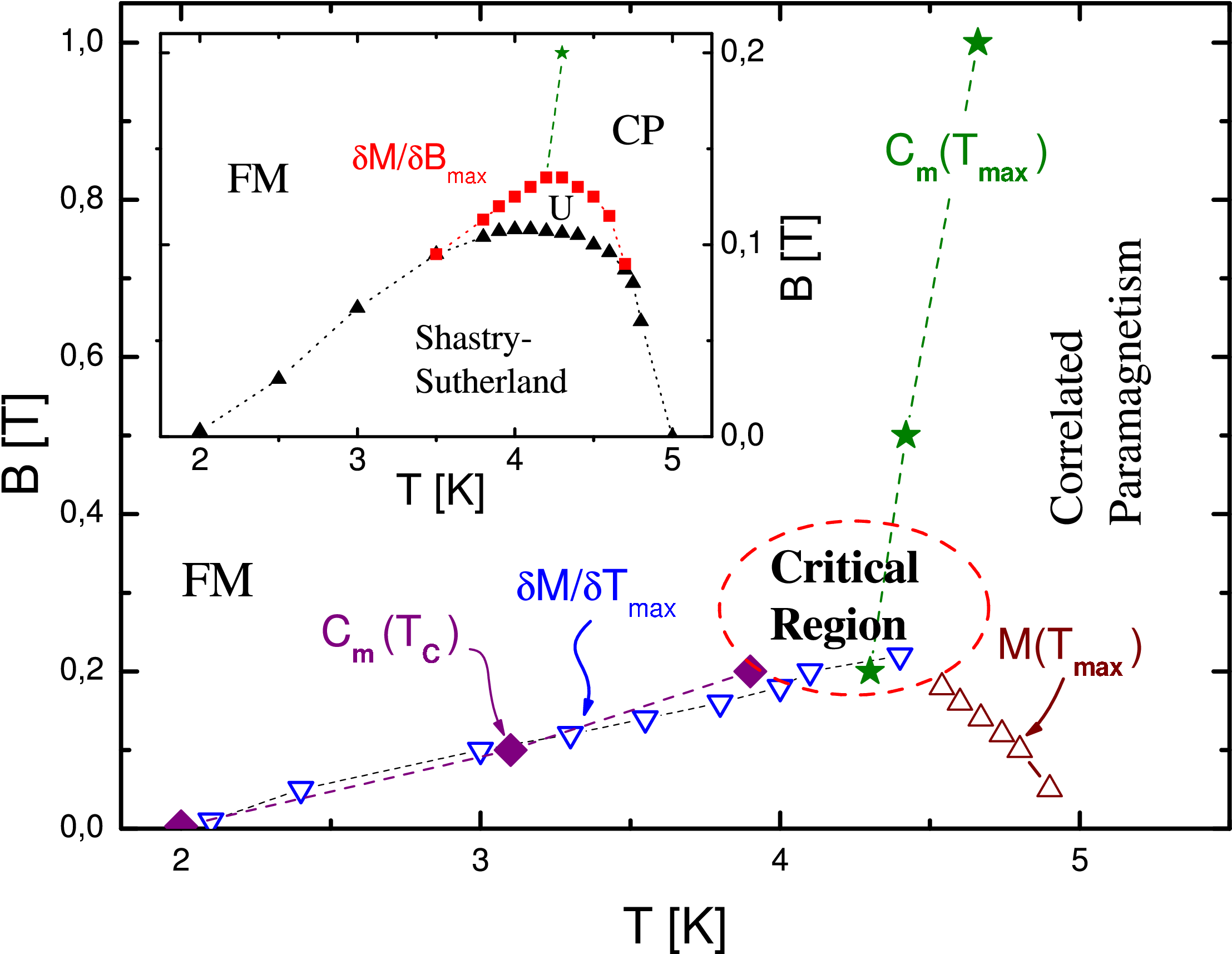}
%{/home/jsereni/papers/publicaz02/pdnial3/textos/F1latpar}
\end{center}
\caption{(Color online) Field vs temperature phase diagram
determined by the temperature of the maximum $\partial M/\partial
T <0$ slope and the $C_m(T_C)$ jump and the cusp of $M(T_M)$.
Inset: detailed of the critical region using the criterion of the
maximum slope of $\partial M/\partial B\mid_{max}$, showing the
site 'U' of the observed structure (see the text). Error bar for
the U-region boundary determination is $\pm 7$\,mT} \label{F6}
\end{figure}

The obtained magnetic phase diagram is presented in Fig.\ref{F6}.
There, the field driven transition between intermediate and FM
phases is drawn according to the temperature of the maximum slope
$\partial M/\partial T\mid_{max}$, which is in agreement with
$C_m(T)$ measurements. Since $T_M$ is practically not affected by
field, both phase boundaries join at a critical point, at
$T_{cr}=(4.2\pm 0.3$)K for $B_{cr}=(0.12\pm 0.02$)T.

In the inset of Fig.~\ref{F6} we show a detail of the critical
region extracted from the maximum slopes of the $M/\partial
B\mid_T(B)$ derivatives depicted in Fig.\ref{F5}b. As mentioned
before, the structure observed in the $\partial M/\partial
B\mid_T$ as a function of field may correspond to one of the steps
predicted in the $M(B)$ dependence, which in this case corresponds
to $\approx 1/8$ of the saturated moment $M_S$. The weakness of
this effect can be attributed to the poly-crystalline character of
the samples.

As proposed in Ref.\cite{Ce2Pd2Sn} for $B=0$, Ce nearest-neighbors
form FM pairs with effective spin $S_{eff}=1$ which define a
simple square lattice lying parallel to the basal plane. A ShSu
phase builds up from those dimers at $T=T_M$, which interact
antiferromagnetically among them. From the present results we
observe that dimers keep forming under moderate external field.
However, the fact that the $C_m(T_C)$ jump and the height of the
$\partial M/\partial B\mid_T$ maxima decrease as $T_C(B)\to T_M$
indicates that the degrees of freedom involved in the formation of
the intermediate phase are progressively reduced by increasing
magnetic field.

\section{Conclusions}

The field dependent magnetic phase diagram was established for
Ce$_2$Pd$_2$Sn. The application of magnetic field confirms that
the transition at $T=T_M$ cannot be regarded as a canonical AF
transition since $T_M$ is practically not affected by fields of
moderate intensity. On the contrary, the low temperature FM phase
is favored by field since the AF interaction among dimers in the
intermediate phase is progressively suppressed.

The instability of the intermediate Shasry-Sutherland phase in
this compound is confirmed by the low value of applied field
$B_{cr}\approx 0.12$\,T required for its suppression. Around the
critical region, a weak structure is observed in the maximum of
$\partial M/\partial B(B)$ dependence at $\approx 1/8 M_S$, which
may be an expected sign for the mentioned phase. In any case,
further investigations are required in order to elucidate whether
this anomalous feature, extending between $3.9< T< 4.6$\,K and
$0.10 < B < 0.14$\,T, confirms the presence of a ShSu phase or it
is due to some other effect.

Other studies are in progress in this family of compounds with the
aim to tune this critical point by doping Ce ligands or alloying
them out of stoichiometry taking profit of the extended range of
solubility of these compounds \cite{Mauro}

\section*{Acknowledgments}
We acknowledge J. Luzuriaga and K. Ogando for their contribution
to magnetic measurements. This work was partially supported by
PIP-6016 (CONICET) and Secyt-UNC project 6/C256. J.G.S. and M.G.B.
are members of CONICET and Instituto Balseiro (UN Cuyo) of
Argentina.

\end{document}